# Tuning electronic and optical properties of free-standing Sn$_2$Bi monolayer stabilized by hydrogenation

Mohammad Ali Mohebpour[1], Sahar Izadi Vishkayi[1], Meysam Bagheri Tagani[1, *]

[1]Computational Nanophysics Laboratory (CNL), Department of Physics, University of Guilan, PO Box 41335-1914, Rasht, Iran

*Corresponding Author. Email: m_bagheri@guilan.ac.ir

## ABSTRACT

In this study, we systematically investigated the structural, mechanical, electronic and optical properties of Sn$_2$Bi monolayer, a sheet experimentally synthesized recently [PRL, 121, 126801 (2018)] which has been hydrogenated (Sn$_2$BiH$_2$) to stabilize free-standing form using density functional theory (DFT). For tuning the band structure and electronic properties, the mechanical strain and electric field are used. Our investigations show that in this free-standing sample, there are electron flat bands and free hole bands like the recently synthesized sample on silicon wafer, which provide the possibility of having strongly localized electrons and free holes with high mobility. Also, the band gap of Sn$_2$BiH$_2$ monolayer has experienced a growth of 80% compared with the experimental sample. The relevant results to strain suggest that the band gap can be properly manipulated by biaxial strain (-13% to +21%) within a range from 0.2 to 1.6 eV. It should be mentioned that the stability and flexibility of the corresponding monolayer under tensile and compressive strain are due to the strong σ bonds between atoms. We also realized the strain can cause indirect-direct transition in the band gap. Furthermore, our optical findings indicate that the Sn$_2$BiH$_2$ monolayer has almost metallic properties in a specific range of UV spectrum and it is transparent in the IR and visible spectrum of electromagnetic radiation. All these tunable properties and nontrivial features portend that Sn$_2$BiH$_2$ monolayer has great potentials in applications as near-infrared detectors, thermoelectric devices, field-effect transistors, sensors, photocatalysis, energy harvesting, and optoelectronics.

**Keywords:** monolayer, electronic properties, optical properties, biaxial strain, density functional theory.

## Introduction

Two-dimensional (2D) crystals are the shining stars in the horizon of materials science and solid state physics which have unique properties, exceptional features, and promising applications. Over the past few years, the interest in 2D materials has grown dramatically. For instance, graphene due to its high carrier mobility allows the electrons to move freely across the material which causes to have an excellent performance in transistors [1, 2]. However, Achilles heel of the electronic structure of graphene, as well as silicene and germanene, is lack of a band gap [3, 4] that reduces severely their capability in switching the on and off current in transistors. However, this issue can be solved by chemical functionalization, applying external electric field and mechanical strain [5-8].

The wide applicable aspects of graphene have encouraged many researchers to explore and synthesize other 2D compounds like hexagonal boron nitride [9], silicene [10], germanene [11], stanene [12], and phosphorene [13]. There is a reasonable motive for investigating these nanoscale materials. These monolayers demonstrate tunable electrical, optical, catalytic, and electrochemical properties [14]. They bring many tempting promises with themselves for the modern technology of FETs [15], sensors [16], photocatalysis [17], field emission [18], and photonic devices [19, 20].

Along with a large number of advanced experimental researches, theoretical studies can also be considered as a more cost-effective alternative for characterizing and predicting the properties of nanomaterials [21, 22]. Evaluating the electronic properties of nanostructure semiconductors is crucial for improving their efficiency in electronics. Of these, applying mechanical strain and external electric fields have a long history in adjusting the electronic properties of semiconductors [23-27]. Researchers have found that nanostructures preserve their structure under tension, very higher and lower than their bulk lattice constants, which unbelievably improves their intrinsic performance in nanotechnology industries [28]. Hence, one could name many materials which can sustain compressive and tensile strains from -10% to +30% due to their high elasticity [25, 29]. Metal-semiconductor transition after applying a certain strain is another positive consequence of this method [30-32]. The researchers have also gained significant achievements in the band gap engineering by applying external electric field. Indeed, they have been trying to tune the band gap of materials for better specifications [33]. For example, a theoretical study suggests that the sizes of band gap in germanene and silicene have linear behavior under electric field [34] and another survey reports the band gap variation of AlAs/germanene heterostructure as 0.5 eV under positive and negative electric field [35].

Very recently, a 2D binary compound with an empirical formula of $Sn_2Bi$ has been synthesized on a silicon (111) substrate which indicates strong spin-orbit coupling and electron-hole asymmetry [36]. In the electronic structure of this system, electron flat bands and free hole bands have also been observed. And this means that by tuning the Fermi level there can be strongly localized electrons and free holes at the same time in this material which can bring attractive features like ferromagnetism and superconductivity. Besides, this structure is very stable because all Si, Bi, and Sn atoms follow the octet rule. The research on $Sn_2Bi$ free-standing monolayer is still at a very early stage and there is not much information available. Only in one theoretical work, the free-standing form of this monolayer is discussed [37]. They showed that the isolated $Sn_2Bi$ monolayer is a metal and suffers from structural and dynamical instability due to the strong bond between the $Sn_2Bi$ and the silicon substrate after synthetization. However, they completely solved this instability issue by chemical functionalization. Similar to the same behavior has been reported in the graphene-like borophene monolayers, that the hydrogenation and halogenation stabilized it dynamically and even improved its optical and thermal conductivity properties [38-40]. Moreover, among monolayers similar to $Sn_2Bi$ free-standing, one can also mention $Mo_2P$ and $Mo_2C$ monolayers which both are metal and stable and have considerable optical properties [41, 42]. Inspired by these, in this paper, by using DFT we present a systematic analysis of the structural, mechanical, electronic, and optical properties of $Sn_2Bi$ free-standing monolayer which has been stabilized by hydrogenation ($Sn_2BiH_2$). We realized that in this 2D system, there are strongly localized electrons and free holes with high mobility like the experimental sample on silicon. We also found that $Sn_2BiH_2$ monolayer is completely stable and its band gap, effective mass, and carrier mobility can be improved and tuned by applying a biaxial strain. Moreover, the optical properties of this monolayer are affected and well improved by strain. The rest of paper is organized as follows: In the next section, we first look at the computational details, in the following, we will discuss the results and eventually we represent our important findings in the conclusion.

## Computational details

In this paper, we have applied the Spanish package solution, SIESTA [43] for structural and electronic simulations which is based on self-consistent density functional theory (DFT) and standard pseudopotentials. For the exchange-correlation interaction, we have utilized the most widely recognized generalized gradient approximation (GGA) given by Perdew, Burke, and Ernzerhof (PBE). These interactions have additionally been approximated by consideration of the spin-orbit coupling (SOC). Because of the underestimation of the band gap by standard DFT, the screened hybrid functional method in HSE06 level is used. This method can estimate the value of band gap more close to experiments by declining the localization and delocalization errors in DFT [44]. During the entire calculations, the density mesh cut-off and convergence tolerance of energy were chosen 150 and $1.0\times10^{-5}$ Hartree, respectively. The reciprocal space was sampled by a mesh of 8×8×1 k-points (for the GGA and SOGGA calculations), and 4×4×1 k-points (for the HSE calculations) in the Brillouin zone. Besides, in order to avoid the interactions between two adjacent unit cells in a periodic configuration, a vacuum space of 20 Å is taken into account in the z-direction. Optimization of the geometry of 2D monolayer has been performed by minimizing the force and stress tolerance down to 0.001 eV/Å and 0.0001 GPa, respectively. Furthermore, to check the stability of the monolayer, the phonon dispersion spectrum was calculated by using a dynamic matrix via the finite-difference method in which a supercell was made with a repetition of 5×5×1 and a k-point sampling of 4×4×1.

The effective mass of electrons and holes (m*) which are gained from the band structure are calculated according to formula (1). In addition, the cohesive energy ($E_c$) and spin-orbit coupling strength ($\lambda_{so}$) of our monolayer can be given by the formulas (2), (3) as below,

$$m^* = \hbar^2 \left(\frac{d^2E}{dk^2}\right)^{-1} \qquad (1)$$

$$E_{cohesive} = \frac{E_{sheet} - \sum_i n_i \, E_{atom\,i}}{N} \qquad (2)$$

$$\lambda_{so} = E_c^{SOC} - E_c^{GGA} \qquad (3)$$

where ℏ is the Planck constant, k is the wave vector in the reciprocal lattice, $E_{sheet}$ and $E_{atom\,i}$ are the total energy of monolayer and the total energy of isolated atom i respectively, $n_i$ represent the number of atom i in the unit cell, N shows the total number of atoms in the unit cell, and $E_c^{SOC}$ indicates the cohesive energy of monolayer considering spin-orbit coupling.

## Results and discussion

### Structural characteristics of Sn$_2$BiH$_2$ monolayer in the stable state

Sn$_2$BiH$_2$ monolayer in which hexagons are formed by combining Bi atoms with a hexagonal network of Sn atoms, has been passivated by hydrogen as shown in Fig. 1a. The above-mentioned honeycomb configuration has ten atoms (two Bi, four Sn, and four H) in a rhombic unit cell. After the optimization of the structure geometry, the lattice constants are a = b = 7.36 Å. This value has experienced a decline of about 0.33 Å in comparison with the synthesized sample on Si (111) [36]. All the optimized structural parameters for this monolayer have been recorded in Table 1 containing the lattice constants (a, b), buckling height (h), bond length (R), cohesive energy ($E_c$) and band gap ($E_g$). Furthermore, our results have been compared with different 2D monolayers such as bismuthene, stanene, arsenene and antimonene which almost have an analogous hexagonal configuration and were synthesized. As can be seen in all these monolayers, the crystal structures are isotropic along a and b directions. Sn$_2$BiH$_2$ monolayer has the largest lattice constant (a = 7.36 Å). While the bond lengths in this layer are much close to the β-antimonene structure, the band gap is similar to β-arsenene. The buckling height of Sn$_2$BiH$_2$ monolayer is equal to 2.46 Å which is much larger than the rest of monoatomic layers, and this is compatible with the fact that by increasing atomic radius the buckling height increases [45]. By evaluating the amounts of cohesive energies of Sn$_2$BiH$_2$ and β-bismuthene, it is quite evident that our monolayer has higher stability.

| structure | a=b (Å) | h (Å) | R (Å) | $E_c$ (eV/Atom) | $E_g$ PBE/SOC (eV) |
|---|---|---|---|---|---|
| Sn$_2$BiH$_2$ | 7.36 | 2.46 | Sn-Sn: 2.83 Sn-Bi: 2.88 | -2.95 | 1.57 / 1.02 |
| β-bismuthene [45] | 4.33 | 1.73 | 3.04 | -2.44 | 0.56 / 0.46 |
| β-stanene [4] | 4.67 | 0.85 | 2.69 | --- | 0.30 / 0.10 |
| β-arsenene [44] | 3.61 | 1.40 | 2.51 | -4.65 | 1.59 / 1.47 |
| β-antimonene [44] | 4.12 | 1.65 | 2.89 | -4.04 | 1.26 / 0.99 |

**Table 1.** The structural parameters of atomic monolayers of Sn$_2$BiH$_2$, Bi, Sn, As and Sb.

In Fig. 1b the phonon dispersion spectrum of Sn$_2$BiH$_2$ monolayer has been plotted. Due to the existence of six atoms (Bi, Sn) in the unit cell excluding H atoms, there are 18 phonon branches. It is quite obvious that all acoustic and optical modes are in the range of positive frequencies, except for an acoustic mode that has negative frequencies of a very small order about -1.6 cm$^{-1}$ nearby the Γ point which is negligible. The maximum of acoustic and optical vibration frequency for this monolayer (excluding branches relevant to H atoms) are 28 and 181 cm$^{-1}$, respectively. These values become 48 and 135 cm$^{-1}$ for buckled β-Bi [45] and go to 48 and 188 cm$^{-1}$ for buckled β-Sn [46], respectively. Nevertheless, the phonon branches of Sn$_2$BiH$_2$ monolayer are smoother which demonstrate that it has less phonon group velocity than β-Bi and β-Sn sheets. Altogether, this phonon band structure offers that our monolayer is dynamically stable.

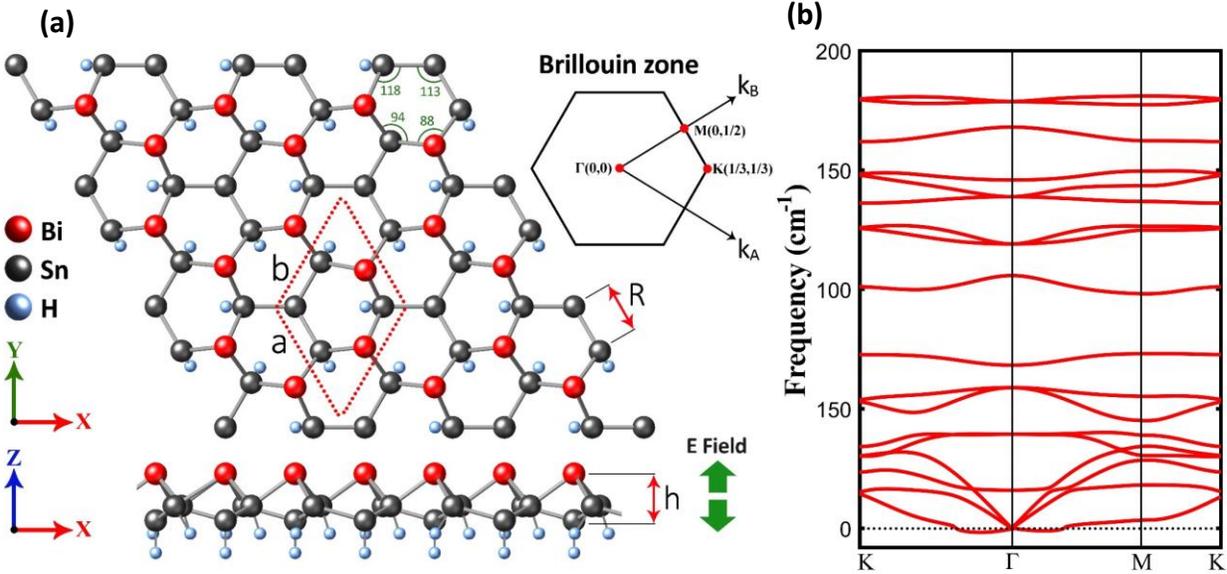

**Figure 1** (a) Top and side view of Sn$_2$BiH$_2$ hexagonal structure together with the rhombic unit cell, lattice constants a=b and brillouin zone. (b) The phonon dispersion spectrum in the GGA level.

## Electronic characteristics of Sn$_2$BiH$_2$ monolayer in the stable state

In Fig. 2 the band structure diagrams of Sn$_2$BiH$_2$ monolayer have been plotted excluding and including SOC and in HSE06 level, respectively. The band structure in GGA level suggests the indirect band gap as 1.57 eV so that the valence band maximum (VBM) and the conduction band minimum (CBM) are located on the Γ and M points, respectively. Of course, this amount of energy gap is reduced to 1.02 eV by considering SOC. Our calculation indicates that Sn$_2$BiH$_2$ monolayer has a negative spin-orbit strength of -0.18 eV which is the reason for the reduction of the band gap in SOC calculations at equilibrium. This quantity is -0.02 and -0.13 for β-As and β-Sb monolayers, respectively. By comparison, one can understand that the negative influence of spin-orbit strength is more significant in Sn$_2$BiH$_2$ monolayer which is due to its relatively heavier atoms. In some of monolayers such as pure germanene, considering the SOC will open the band gap up to 24 meV. In fact, the spin-orbit strength has a positive effect. This effect also increases in the halogenated samples of germanene [47]. It is also clear that the predicted band gap of Sn$_2$BiH$_2$ monolayer is larger than buckled β-Bi, β-Sn and β-Sb and close to β-As, which will be a promising candidate in nanoelectronics and optoelectronics [48]. Further, this value has confronted with an increment of 80% than the band gap of the experimental sample on Si (111) calculated by angle-resolved photoemission spectroscopy (ARPES) method ($E_g = 0.87$ eV). It should be noticed that including the SOC has no effect on the degeneracy of the energy levels in VBM whereas, it removes the degeneracy of VBM in both buckled Bi and Sn monolayers [45, 49]. It can be easily found from Fig. 2b that there are electron flat bands and free hole bands. So, we will be able to have free and strongly localized charge carriers through this 2D monolayer by tuning the Fermi level. In VBM which is located at $E_v = -0.47$ eV, the effective mass of light and heavy holes in two different directions Γ-K and Γ-M are the same as $m^* = -0.26\ m_e$ which is indicative of the energy symmetry and high mobility for free hole gas. The gained $m^*$ is larger than the effective mass of holes in $Sn_2Bi$-Si (111) experimental sample ($m^* = -0.14\ m_e$) [36]. By comparing the band structure of our monolayer with the synthesized sample on silicon (in the presence of SOC), it can be seen that the location of VBM and CBM are the same. The energy levels at the Γ point are degenerate in both systems so that the valence levels near the Fermi energy are like two cones touching each other at the top (the degree of degeneracy is two). Moreover, considering the SOC will create a huge Rashba splitting for the lowest conduction band around the Γ point (Rashba effect is lifting the electron-spin degeneracy because of SOC interactions). This is highly desirable for electricity relevant for spintronic functions. Furthermore, the distance between valence band energy levels at the Γ point is more than that of $Sn_2Bi$-Si (111) experimental sample [36]. In HSE06 level, the band structure exhibits an indirect band gap as 2.01 eV. In this status, the VBM is still located at the Γ point but the CBM has been shifted to the K point (see Fig. 2c).

In order to see the contribution of different orbitals around the Fermi surface in the band structure of $Sn_2BiH_2$ monolayer, we have calculated the density of states (DOS) as shown in Fig. 3a. Due to the high atomic weight of Bi (Z=83) and Sn (Z=50), the energy-dependent hybridization between these atoms and the great influence of SOC on the band gap, the SOC has been taken into account in here. As it turns out, the contribution of p orbitals of both atoms is quite dominant in the valence band maxima. This domination can be seen too in the conduction band minima for Bi atom while for Sn atom, the portion of s and p orbitals are identical according to the electronic configuration of Sn atom in the last shell ($5s^2\ 5p^2$). Our results are in good agreement with this fact that, in most of monoatomic layers with honeycomb structures such as arsenene [26], antimonene [44], and bismuthene [45] where the bonds are of σ kind, the contribution of p orbital is dominant than the others.

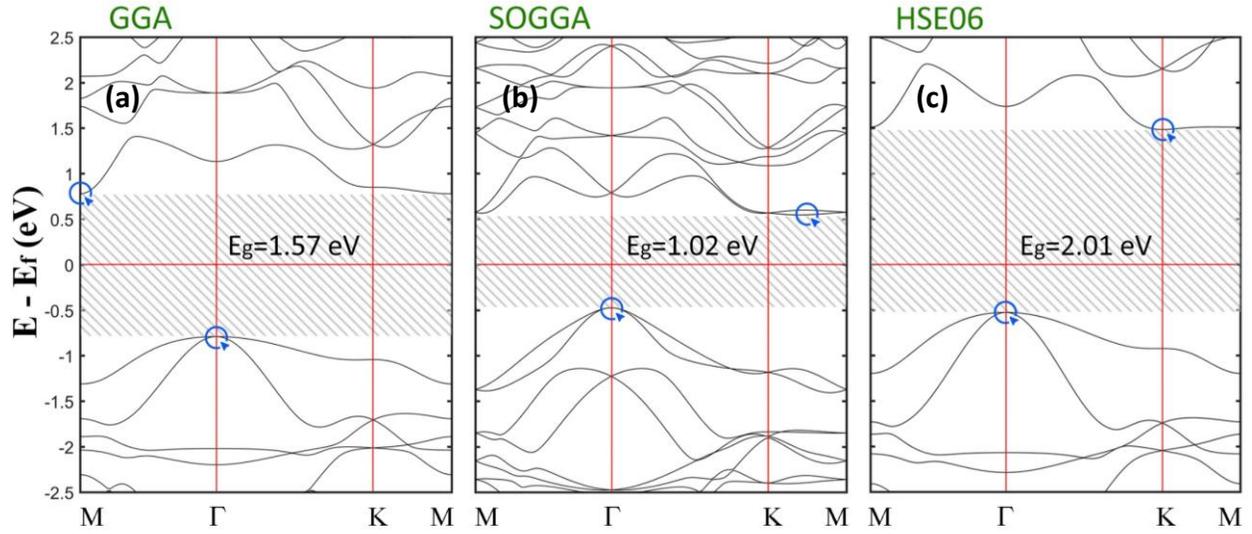

**Figure 2** The band structure diagrams of $Sn_2BiH_2$ monolayer (a) excluding and (b) including the SOC and in (c) HSE06 level. The locations of VBM & CBM have been specified with blue circles.

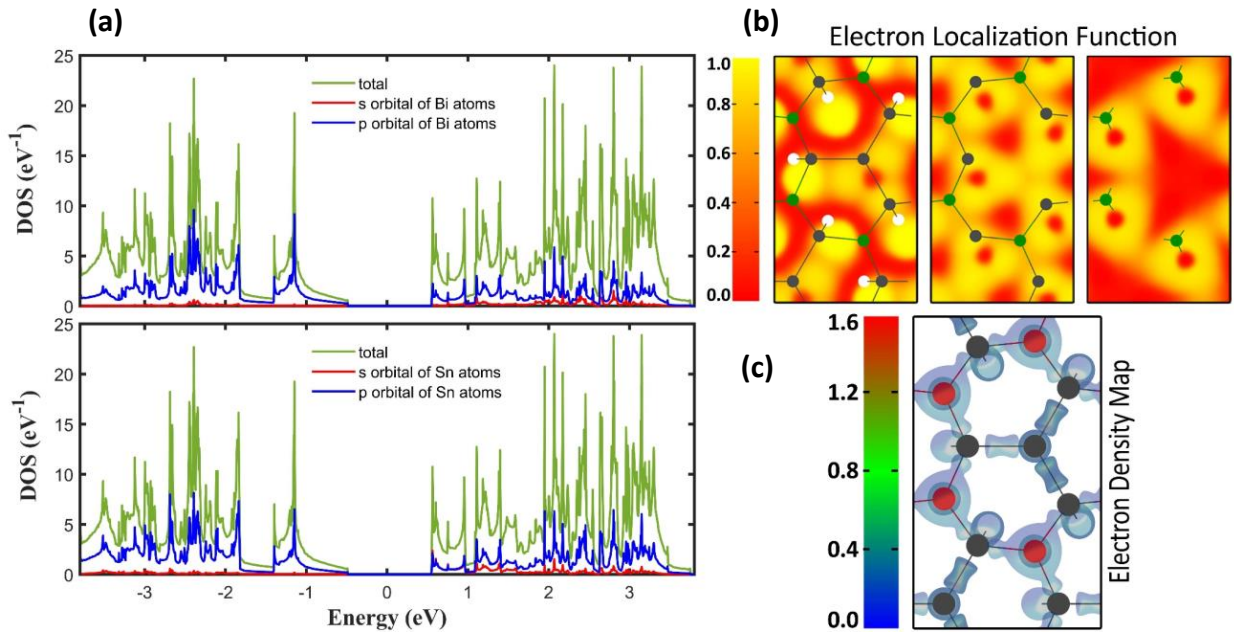

**Figure 3** (a) The density of states of $Sn_2BiH_2$ monolayer. (b) The electron localization function in three different plane in the z direction. The first plane consists of H atoms, the second and third pass through Sn and Bi atoms, respectively. (c) The electron density map in the (001) direction with the isovalue set to 0.26 e.Å$^{-3}$.

To be able to understand the spatial localization of electrons in the neighborhood of atoms, the electron localization function (ELF) has been calculated. The ELF is a location-dependent scaling function that has a range from zero to one, where ELF = 1 represents the highest localization. In Fig. 3b, the ELF profile has been plotted for intended monolayer in three different locations in the (001) direction (The first ELF profile has been depicted at the plane of H atoms, the second at the plane of Sn atoms and the third at the plane of Bi atoms). It is completely obvious that the electrons possess the highest localization near the H atoms while the ELF scale is almost the same in the vicinity of Sn and Bi heavy atoms. It can be understood as well that the hydrogenation has made the Sn$_2$Bi monolayer stable. Because this monolayer has surplus electrons and is prone to donate them which justify its structural instability but after passivation, the electrons are transferred to hydrogen atoms. In Fig. 3c the electron density profile of this monolayer has been drawn in the (001) direction. It is clearly seen that the electron density between two Sn atoms with sp$^3$ hybridization has significant amount, so these atoms form a σ bond. Moreover, the bonds between Bi and Sn atoms are of a σ type but the electrons tend to move towards Bi atoms according to the larger electronegativity of this atom in Pauling scale ($\chi^{Bi}$ = 2.02, $\chi^{Sn}$ = 1.96) [50]. Since the electronic configuration of Bi atom in the last shell is 6s$^2$ 6p$^3$, the electron density around this atom with sp$^2$ hybridization arises only from the p orbital and is dumbbell-shaped. On the other hand, the Mulliken population analysis indicates that the charge is transferred from the Bi and Sn atoms to the H atoms ($\delta q^H$ = +0.13e, $\delta q^{Bi}$ = -0.15e, $\delta q^{Sn}$ = -0.05e). This can justify the electron density and the high electron localization around the hydrogen atoms.

## Electronic characteristics of Sn$_2$BiH$_2$ monolayer under biaxial strain

We are truly informed that mechanical strain plays an impressive role in tuning and improving the electronic properties of 2D materials. Since the Sn$_2$BiH$_2$ monolayer is mechanically isotropic in the xy-plane, we disregard the strain in the separate directions and only apply it in the form of biaxial. It should be noted that it is applied as tension (positive strain) and compression (negative strain). In Fig. 4a the stress-strain change curve has been plotted. Because the Sn$_2$BiH$_2$ is a monolayer, the stress values are only multiplied by the length of the unit cell (20 Å) in z-direction to have the unit of N/m. It is simple to figure out that as the strain increases, the stress enhances nonlinearly. According to our calculation, the ideal strength which is defined based on the minimum and maximum of the stress, is -4.9 and 2.9 N/m. The obtained values for our monolayer are almost similar to silicene (5 N/m) [51] and phosphorene (4 N/m) [25] monolayers under strain in zigzag and armchair directions, respectively. The mentioned ideal strengths appear under -13% and +21% critical strains. It is worth noting that the stability of Sn$_2$BiH$_2$ monolayer up to this range is due to the strong σ bonds between atoms. The reported critical strains for borophene, silicene, stanene and even antimonene monolayers are about +12%, +20%, +17%, and +15%, respectively [51-54] which are less than ours and showing the acceptable mechanical stability and flexibility of Sn$_2$BiH$_2$ monolayer. These features together make Sn$_2$BiH$_2$ monolayer a competitive candidate in 2D flexible nanoelectronics for manufacturing of touch screens, smart watches, and transistors [55]. On the other side, in some honeycomb structures like graphene and phosphorene, the critical strains have been reported as +19% and +27% in the zigzag and +26% and +30% in the armchair direction, respectively [25, 56] which have more ideal strength than that of our monolayer. In Fig. 4b the band gap curves as a function of strain in either presence or absence of the SOC and in the HSE6 level have been depicted. As it is obvious, the band gap change behavior is almost analogous and completely tunable in all three approximations. By increasing the tensile strain, the band gap reaches its own peak ($E_g^{GGA}$ =1.62, $E_g^{SOGGA}$=1.05, $E_g^{HSE}$=2.05 eV) and then it experiences a linear reduction. Also, as the compression rate enhances, the band gap falls more sharply, although in SOGGA level and within the range from -08% to +02%, the band gap modification is not clear and remains the same. The observed behavior is very similar to arsenene and phosphorene monolayers under biaxial and uniaxial strain in the zigzag direction, respectively. But in the case of arsenene and phosphorene, the band gaps are closed from a range on [57, 58]. It is very interesting that $Sn_2BiH_2$ monolayer maintains semiconducting feature even under high compression or tension. Overall, the band gap reduction is arising from the downward movement of the CBM and the upward movement of VBM simultaneously. To better realize the band gap behavior under strain, the band structures for each strain have been depicted in the Online Resource Fig. S1. In the GGA level, an indirect-direct band gap transition emerges under +05% strain, and the direct band gap remains still to +18%. This transition can be found between -11% and -13% as well. There is no indirect-direct gap transition in the presence of SOC. In the HSE06 level, the band gap is only indirect in the range from -04% to +02% and turns to direct in the rest of strains. It should be mentioned that the strain also modifies the effective mass of the charge carriers. In particular, those critical strains in which indirect-direct gap transition occurs. In the GGA level, under +05% strain for instance, the effective mass of holes in the VBM jumps to m* = -0.87 $m_e$ while the effective mass of electrons in the CBM drops to m* = 0.18 $m_e$ (in the armchair direction). And that means one can have charge carriers with high mobility and conductivity by applying the strain. The results are in line with similar theoretical works [58].

The buckling height (h) is an effective factor for describing the corrugation of 2D materials [52, 59]. To see this, in Fig. 5 the value of buckling height of $Sn_2BiH_2$ monolayer as a function of strain has been drawn. It can be clearly seen that by applying the strain, as the tension increases, the buckling height decreases gradually and this trend continues to the failure strain (+21%) which effectively flatten the wrinkles of the monolayer surface. The buckling height of $Sn_2BiH_2$ is 1.73 Å under +10% strain that is equal to the β-Bi monolayer [45], but because of the higher cohesive energy, $Sn_2BiH_2$ has more mechanical stability. In opposite, when the strain increases as compression, the buckling height rises with the same slope. Increasing the buckling height also make changes in the electron density and enhance the spin-orbit strength to +0.26 eV (under -13% strain). For this reason, the band gap of $Sn_2BiH_2$ monolayer in the SOGGA level ($E_g$ = 0.61 eV) is larger than that in the GGA level ($E_g$ = 0.23 eV) under -13% strain. Applying a tensile strain not only flattens the buckled and puckered monolayers but also increases the bond length (R) as it is observed in Fig. 5. It is clear that as the tension increases, both bond lengths (Sn-Sn and Sn-Bi) undergo a monotonous and near-linear growth (~0.26 Å). Although under compressive strain, the Sn-Sn bonds decline twice as the Sn-Bi bonds ($\Delta R_{Sn-Sn}$= 0.12). This gradual increment and decrement cause the weakening of the strong σ bonds between atoms which somewhat reduces the mechanical stability of the structure.

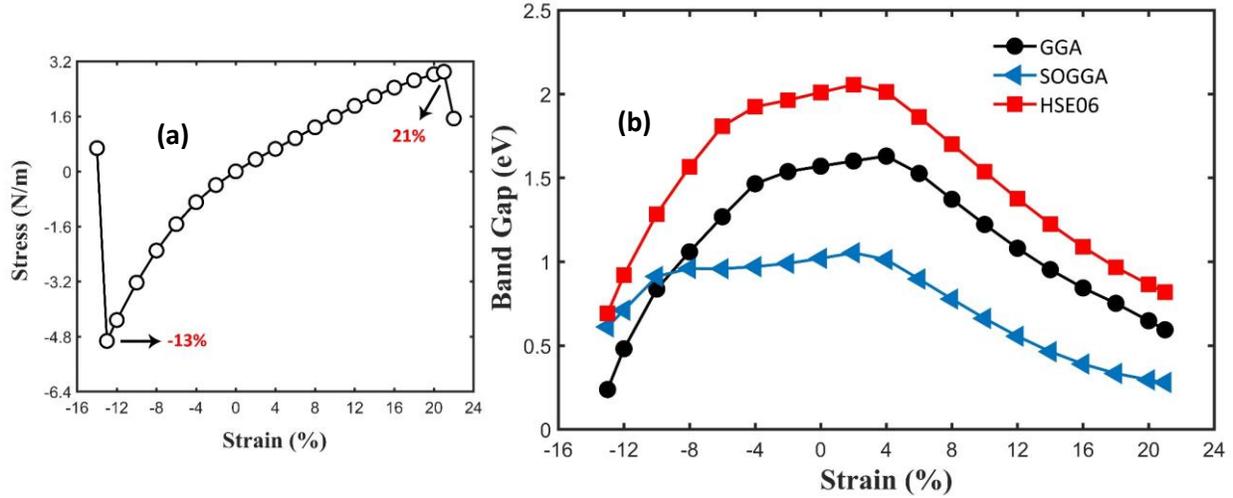

**Figure 4** (a) The stress-strain curve of $Sn_2BiH_2$ monolayer with the failure strains given. (b) The variation of band gap as a function of strain in either presence or absence of the SOC and in the HSE06 level.

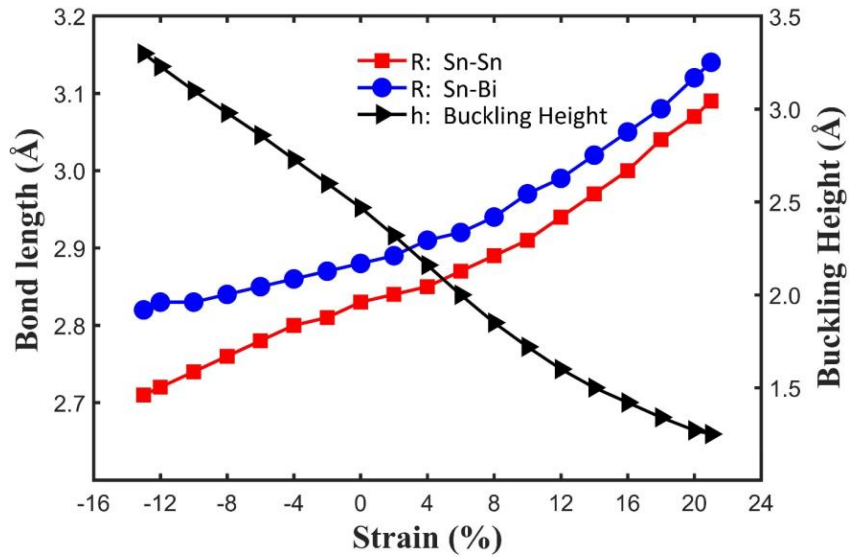

**Figure 5.** The variation of bond length and buckling height of $Sn_2BiH_2$ monolayer as a function of strain.

## Band gap engineering using an external electric field

A large number of theoretical and experimental studies have demonstrated that the external electric field is one of the powerful instruments for exploring the electrical properties of 2D materials [35, 60-62]. Therefore, in this section, we intend to investigate the impact of E-field on the electronic properties of $Sn_2BiH_2$ monolayer. In Fig. 6, the band gap variation as a function of E-field has been plotted in the GGA level. These calculations have also been developed for two compressive (-02%) and tensile (+05%) strains. It is clearly seen that although the band gap behavior is not showing a high sensitivity to the E-field, it slightly changes. The band gap of pristine and -02% strained sheet increase monotonically (~40 meV) by increasing the positive E-field. Similar to these linear behaviors with a little more sensitivity is observed in the semi-metallic single-layer buckled silicene and germanene in which the E-field will open the band gap as 100 meV [34]. For +05% strained sheet, the band gap response to the E-field shows a descending variation. This is also seen in the buckled bismuthene under the same E-field [45]. In the pristine and -02% strained sheet, the band gap type is independent of the E-field but for +05% strained sheet, a direct-indirect transition appears under fields stronger than -0.5 V/Å. As long as the E-field is applied in the negative direction (upwards), we expect that the charge transfer is facilitated from Sn and Bi atoms to H atoms. This transfer is confirmed by the Mulliken population analysis of $Sn_2BiH_2$ monolayer under the field -0.7 V/Å ($\delta q^H$ = +0.18e, $\delta q^{Bi}$ = -0.31e, $\delta q^{Sn}$ = -0.15e). Higher intralayer charge transfer means more charge separation which will end in stronger bonds and higher stability. Therefore, the $Sn_2BiH_2$ monolayer is in a more stable situation under a negative E-field so that this can be easily seen from the cohesive energy curve as a function of E-field in the Online Resource Fig. S2.

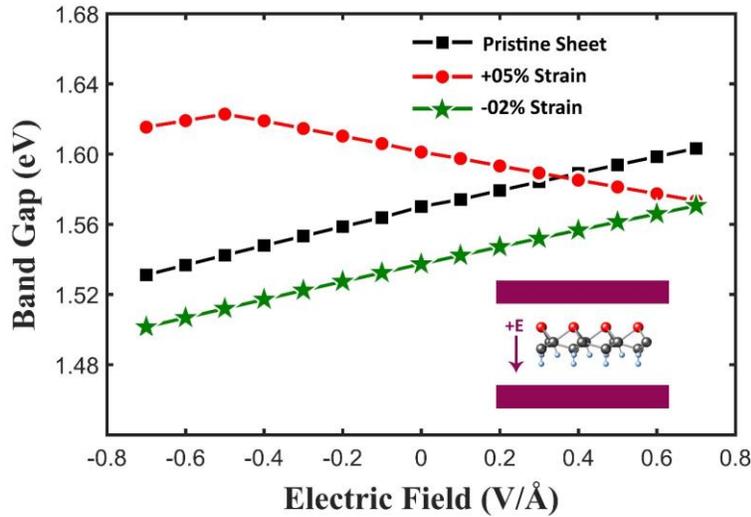

**Figure 6** The band gap variation with E-field for the pristine, +05% and -02% strained sheet. The positive direction of E-field has been specified in the scheme which is from the higher to lower potential.

## Optical characteristics of $Sn_2BiH_2$ monolayer

To investigate the optical properties of 2D crystals, we need a complex dielectric function. A function that represents the reaction of a material to the applied electromagnetic field, and plays a vital role in determination of the features and applications of the emerging 2D materials [35, 63-65]. In this part, applying the dielectric function with parallel and perpendicular polarization vectors to the surface of the monolayer, we discuss the optical properties of $Sn_2BiH_2$ monolayer for the pristine form, +02% and +06% deformed structures (see Fig. 7). There are convincing reasons for choosing each of the two specific strain and they are: The band gap value of +02% strained monolayer is larger than that in the pristine and the band gap type of the +6% strained monolayer is direct.

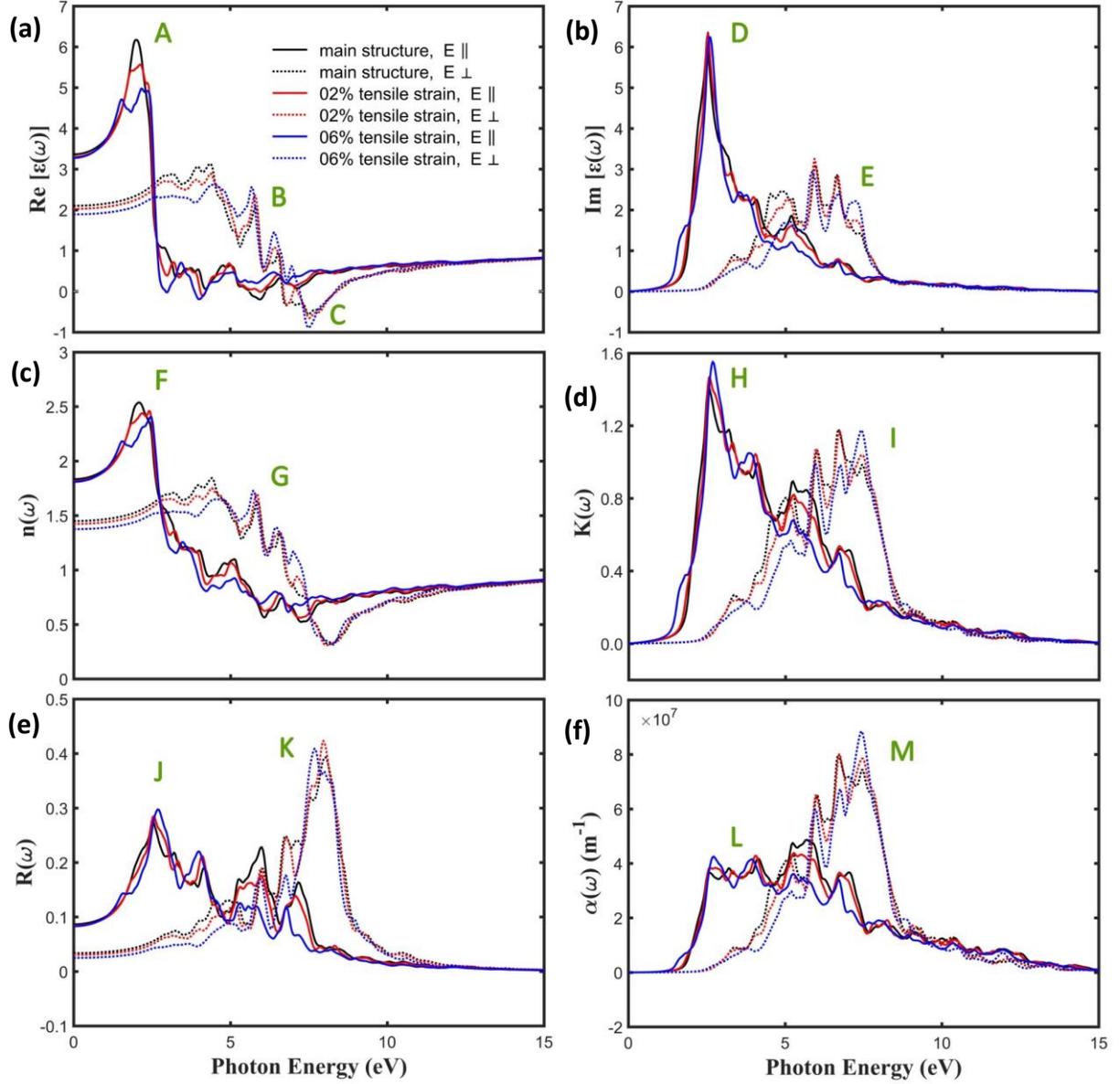

**Figure 7** The optical results of $Sn_2BiH_2$ monolayer as a function of photon energy. (a) The real and (b) imaginary parts of the dielectric function, (c) refractive index, (d) extinction coefficient, (e) reflectivity, and (f) the absorption coefficient for the pristine (black), +02% (red) and +06% (blue) deformed structures. All the optical results have been calculated with in-plane (E ∥) and out-of-plane (E ⊥) light polarizations since the $Sn_2BiH_2$ monolayer is isotropic in the xy-plane and the results are similar along x and y direction. These calculations are done in the GGA level using 21×21×1 k-point sampling.

It should be mentioned that for investigation of the optical properties such as the refractive index n(ω), extinction coefficient k(ω), reflectivity R(ω), and the absorption coefficient α(ω), we calculate the complex dielectric function ε(ω) in the framework of time-dependent DFT method in the independent particle approximation. This calculation is specified at the first-order perturbation theory by means of self-consistent ground-state DFT energies and eigenfunctions that are plugged into the dipolar transition matrix elements. In this regard, at first the Kubo-Greenwood formula [66] is used to calculate the susceptibility tensor $\chi_{ij}(\omega)$,

$$\chi_{ij}(\omega) = -\frac{e^2\hbar^4}{m^2\varepsilon_0 V \omega^2} \sum_{nm} \frac{f(E_m) - f(E_n)}{E_{nm} - \hbar\omega - i\Gamma} \pi^i_{nm} \pi^j_{mn} \tag{4}$$

where $\pi^i_{nm}$ is the i-th component of the dipole matrix element between state n and m, $\varepsilon_0$ the vacuum dielectric constant, V the volume, ω the frequency of electro-magnetic wave, Γ the broadening, and f the Fermi function. Then, through the equation (5) and the susceptibility tensor, the dielectric tensor can be easily achieved. This tensor can also be written as equation (6),

$$\varepsilon(\omega) = 1 + \chi(\omega) \tag{5}$$

$$\varepsilon(\omega) = \varepsilon_1(\omega) + i\varepsilon_2(\omega) \tag{6}$$

where $\varepsilon_1(\omega)$ and $\varepsilon_2(\omega)$ are the real and imaginary parts of the complex dielectric function, respectively. Other optical quantities are obtained using the real and imaginary parts of this function as below:

$$n(\omega) = \frac{1}{\sqrt{2}} \sqrt{\sqrt{\varepsilon_1^2(\omega) + \varepsilon_2^2(\omega)} + \varepsilon_1(\omega)} \tag{7}$$

$$K(\omega) = \frac{1}{\sqrt{2}} \sqrt{\sqrt{\varepsilon_1^2(\omega) + \varepsilon_2^2(\omega)} - \varepsilon_1(\omega)} \tag{8}$$

The reflectivity is acquired directly from Fresnel's formula, and the absorption coefficient is related to the extinction coefficient through equation (10) in which c represents the speed of light in vacuum.

$$R(\omega) = \frac{[1 - n(\omega)]^2 + K(\omega)^2}{[1 + n(\omega)]^2 + K(\omega)^2} \tag{9}$$

$$\alpha(\omega) = \frac{2\omega K(\omega)}{c} \tag{10}$$

In Fig. 7a, b the real and imaginary parts of dielectric function have been depicted in two parallel (E ∥) and perpendicular (E ⊥) polarizations. It is clearly seen that the dielectric function is completely anisotropic in these directions so that the values and the oscillations are more intense in the parallel polarization. The real part of dielectric function shows a peak indicated by A at 2.0 eV for E ∥ while it presents three peaks indicated by B in the energy range from 4.0 to 7.0 eV for E ⊥. This function displays negative values in the range from 5.6 to 6.1 eV as well as from 6.7 to 8.4 eV for E ∥ and E ⊥, respectively which demonstrate the metallic properties of $Sn_2BiH_2$ monolayer in this range of the UV spectrum. In the presence of strain and by increasing the tension, the peak height indicated by A decreases and shows a blue shift but such behavior in the peaks indicated by B is not observed. The static dielectric constant $\varepsilon_1(0)$ of this monolayer is equal to 3.4 for E ∥ which is larger than that in β-As and β-Sb monolayers and has a higher polarization [44]. For the imaginary part of dielectric function, there is a peak indicated by D in the visible light spectrum at 2.48 eV for E ∥ whereas there are two peaks indicated by E in the UV spectrum at 5.96 and 6.64 eV for E ⊥ (see Fig. 7b). As we know, these peaks are caused by the absorption of incident photons and direct interband transitions by the valence band electrons [67]. In the Online Resource Fig. S3, by comparing the energy difference of levels higher and lower than the Fermi surface with the energy of peaks in the imaginary part of dielectric function, we have determined the probable location of the interband transitions. It is also evident from the Fig. 7b that the location and intensity of the peaks modify slightly and do not show high sensitivity with an increase in the strain.

In Fig. 7c the refractive index of $Sn_2BiH_2$ as a function of incident photon energy has been plotted. The refractive index maximum n = 2.5 indicated by F is seen in the visible spectrum at 2.1 eV for E ∥ nearly the energy in which the real part of dielectric function shows a peak while a maximum of n = 1.8 is observed in the UV spectrum at 4.4 eV for E ⊥. This indicates that the scattering of light beams by $Sn_2BiH_2$ monolayer strongly depends on the incident light polarization. The static refractive index n(0) of this monolayer is equal to 1.8 for E ∥ that is larger than β-Sb ($n_0$ = 1.5) and smaller than α-Sb ($n_0$ = 2.3) [44]. The extinction coefficient is inconsiderable in the visible light (1.5-3 eV) for E ⊥ whereas increases dramatically for E ∥ (see Fig. 7d). The extinction coefficient maximum indicated by H is located in the visible spectrum at 2.5 eV for E ∥ nearly the energy in which the imaginary part of dielectric function has peaked.

Fig. 7e represents the reflectivity curves as a function of photon energy. As can be seen, the highest reflection for E ∥ occurs in the visible region while for E ⊥ no significant reflection is observed in the entire energy range less than 5 eV. Similar to this behavior can be seen in the absorption curves of $Sn_2BiH_2$ monolayer (see Fig. 7f). Although the absorption rate is quite negligible in the energy range less than 5 eV for E ⊥ that makes the monolayer transparent, it reaches its strongest limit in the UV region (labeled with M). The zero absorption efficiency in the IR region will mean that, $Sn_2BiH_2$ monolayer does not attract the heat and can be used as a beam splitter or optical fibers. There is less absorption for E ∥ rather that E ⊥ and its edge is formed by visible light (labeled with L).

This article deals with the theoretical study of Sn$_2$Bi monolayer stabilized by hydrogenation. However, the proposed sample can be synthesized experimentally by chemical vapor deposition (CVD) method and liquid exfoliation. First of all, to prepare the β-Bi/Si(111), about 1.2 monolayer (ML) of Bi atoms should be deposited on a silicon (111) surface at room temperature and subsequently annealed at 670 K [68]. Then, by deposition of Sn atoms on the β-Bi surface at 470 K, the honeycomb Sn$_2$Bi is formed [36]. The fabricated structure is on the silicon substrate but the free-standing form of Sn$_2$Bi can be achieved by the liquid exfoliation of layered structures [69]. To stabilize the Sn$_2$Bi free-standing monolayer, it should be hydrogenated. Hydrogen plasma treatment is a known method for modification of the surface of graphene monolayer [70]. In this method, first, the hydrogen plasma is generated in a separate gun with a microwave power of 180W at 2.5 GHz. Then, the prepared monolayer by CVD is placed 10 cm away from the end of the gun while the flow rate of hydrogen is 25 SCCM and the working pressure is 10 mtorr exactly similar to the hydrogenation of graphene surface [70].

## Conclusion

In summary, we investigated the structural, electronic and optical properties of Sn$_2$BiH$_2$ free-standing monolayer using the first principle calculations. Our results show that in this free-standing monolayer there are strongly localized electrons as well as free holes with high mobility exactly similar to the synthesized sample on silicon (111). Also, it is suggested that Sn$_2$BiH$_2$ has an indirect band gap that the band gap type and value, the effective mass and as a result the mobility are tuned by applying a mechanical strain. This monolayer maintains its mechanical stability under the applied strain up to high values (-13% to +21%) because of the existence of σ bonds between atoms, which indicates its substantial flexibility. For the dielectric function, perfect isotropy is observed along the x and y directions which is due to the isotropic crystal structure in xy-plane. When the polarization is E ⊥, the absorption is extremely low and the reflection is negligible in the energy range less than 5 eV which is indicative of the high transparency of Sn$_2$BiH$_2$ free-standing monolayer for the IR and visible spectrum. All these significant features together convert Sn$_2$BiH$_2$ free-standing monolayer as a qualified candidate in nanoelectronics and optoelectronics for manufacturing of the flexible smart systems.